\documentclass{aa}
\usepackage{graphicx}
\usepackage{latexsym}
\newcommand{\bra}{Br$\alpha$}
\newcommand{\brb}{Br$\beta$}

\newcommand{\pfa}{Pf$\alpha$}

\newcommand{\pfg}{Pf$\gamma$}
\newcommand{\pfd}{Pf$\delta$}
\newcommand{\pff}{Pf(11-5)}
\newcommand{\pfh}{Pf(13-5)}
\newcommand{\hu}{Hu(14-6)}
\newcommand{\heii}{\ion{He}{ii}}
\newcommand{\hei}{\ion{He}{i}}
\newcommand{\hi}{\ion{H}{i}}
\newcommand{\oi}{\ion{O}{i}}
\newcommand{\feii}{\ion{Fe}{ii}}
\newcommand{\mgii}{\ion{Mg}{ii}}
\newcommand{\siiv}{\ion{Si}{iv}}
\newcommand{\mum}{$\mu$m}
\newcommand{\msunyr}{M$_{\odot}$~yr$^{-1}$}
\newcommand{\micron}{\ensuremath{\mu {\rm m}}}
\newcommand{\lam}{$\lambda$}
\newcommand{\kmsec}{\mbox{km~s$^{-1}$}}
  
\newcommand{\mdot}{\ensuremath{\dot{M}}}              
\newcommand{\rstar}{\ensuremath{\mathrm{R}_{\star}}}  
\newcommand{\teff}{\ensuremath{\mathrm{T}_{\rm eff}}} 
\newcommand{\vinf}{\ensuremath{\mathrm{v}_{\infty}}} 
\newcommand{\weq}{\ensuremath{\mathrm{W}_{\rm eq}}}   
\begin{document}

\thesaurus{07       
          (02.12.2; 
	   04.03.1; 
	   08.05.1; 
	   08.06.3; 
           13.09.6)}
\title{An atlas of 2.4 to 4.1 \mum\ ISO/SWS spectra of early-type stars 
\thanks{Based on observations with ISO, an ESA
project with instruments funded by ESA Member States (especially
the PI countries: France, Germany, the Netherlands and the United
Kingdom) and with the participation of ISAS and NASA}}
\titlerunning{Diagnostics from near-infrared ISO/SWS spectra of early-type stars}
\author{A. Lenorzer\inst{1}, B. Vandenbussche\inst{2}, P. Morris\inst{3}, 
A. de Koter\inst{1}, 
T.R. Geballe\inst{4}, L.B.F.M. Waters\inst{1,2}, S. Hony\inst{1} \and 
L. Kaper\inst{1}}
\authorrunning{A. Lenorzer et al.}
\offprints{A. Lenorzer (lenorzer@astro.uva.nl)} 
\institute{Sterrenkundig Instituut ``Anton Pannekoek'', Kruislaan 403, NL-1098 
SJ Amsterdam \and Instituut voor Sterrenkunde, K.U. Leuven, Celestijnenlaan 
200B, B-3001 Heverlee \and
SIRTF Science Center / IPAC, California Institute of Technology, M/S 100-22, 
1200 E. California Blvd., Pasadena, CA 91125  USA
\and Gemini Observatory, 670 N. A'ohoku Place, Hilo, HI 96720 USA}
\date{October 2001}

\maketitle

\begin{abstract}

     We present an atlas of spectra of O- and B-type stars,
     obtained with the Short Wavelength
     Spectrometer (SWS) during the Post-Helium program of the Infrared
     Space Observatory (ISO). This program is
     aimed at extending the Morgan \& Keenan classification scheme into
     the near-infrared. Later type stars will be discussed in a seperate
     publication.
     The observations consist of 57 SWS Post-Helium spectra from
     2.4 to 4.1 \micron, supplemented with 10 spectra acquired
     during the nominal mission with a similar observational setting.
     For B-type stars, this sample provides ample spectral converage
     in terms of subtype and luminosity class. For O-type stars,
     the ISO sample is coarse and therefore is complemented with 8
     UKIRT L'-band observations.
     In terms of the presence of diagnostic lines, the L'-band is
     likely the most promising of the near-infrared atmospheric
     windows for the study of the physical properties of B stars. 
     Specifically, this wavelength interval contains the \bra, \pfg,
     and other Pfund lines which are probes of spectral type, luminosity
     class and  mass loss. 
     Here, we present simple empirical methods based on the lines
     present in the 2.4 to 4.1 \mum\ interval that allow
     the determination of
     {\em i)}   the spectral type of B dwarfs and giants to within 
                two subtypes;
    {\em ii)}  the luminosity class of B stars to within two classes;
    {\em iii)} the mass-loss rate of O stars and B supergiants to
                within 0.25 dex.


\end{abstract}

\section{Introduction}

\begin{table*}[!ht]
 \begin{center}
 \caption{The 12 O-type stars and one Wolf-Rayet star observed during the 
 ISO/SWS Post-Helium mission, supplemented with 7 O stars observed with 
 CGS4/UKIRT. The spectrum averaged signal-to-noise ratio (S/N) is listed 
 in the last column.}
 \label{OB1}  
 \begin{tabular}{lclcccr}
 \hline	    
 \hline	    
 Star&Name&Spectral&Spect. Type             &ISO Observation&Instrument&S/N\\
     &    &Type    &Reference$^{\mathrm{a}}$&   Number      &          &   \\ 
 \hline	    
 \object{HD~46223  }&NGC 2244 203  &O4V((f))    &W72&            &UKIRT    &220\\ 
 \object{HD~190429A}&              &O4If+       &W73&89300401    &ISO/PHe  & 15\\ 
 \object{HD~46150  }&NGC 2244 122  &O5V(f)      &W72&            &UKIRT    &150\\    
 \object{HD~199579 }&HR8023        &O6V((f))    &W73&89300301    &ISO/PHe  & 20\\
 \object{HD~206267 }&HR8281        &O6.5V((f))  &W73&90001601    &ISO/PHe  & 30\\  
 \object{HD~47839  }&15 Mon        &O7V((f))    &W72&            &UKIRT    & 70\\   
 \object{HD~24912  }&$\zeta$ Per   &O7.5III((f))&W73&            &UKIRT    &125\\
 \object{HD~188001 }&QZ Sge        &O7.5Iaf     &W72&90000801    &ISO/PHe  & 10\\ 
 \object{HD~36861  }&$\lambda$ Ori A&O8III((f)) &W72&            &UKIRT    &180\\
 \object{HD~209481 }&LZ Cep        &O9V         &W73&90001701    &ISO/PHe  & 25\\ 
 \object{HD~193322 }&HR7767        &O9V((n))    &W72&88201401    &ISO/PHe  & 15\\ 
 \object{HD~37043  }&$\iota$ Ori   &O9III       &W72&            &UKIRT    &140\\
 \object{HD~207198 }&HR8327        &O9Ib-II     &W72&88502001    &ISO/PHe  & 25\\ 
 \object{HD~38666  }&$\mu$ Col     &O9.5V       &W73&90701901    &ISO/PHe  & 10\\ 
 \object{HD~37468  }&$\sigma$ Ori  &O9.5V       &C71&            &UKIRT    &165\\
 \object{HD~209975 }&19 Cep        &O9.5Ib      &W72&90001501    &ISO/PHe  & 30\\ 
 \object{HD~188209 }&HR7589        &O9.5Iab     &W72&88000501    &ISO/PHe  & 20\\ 
 \object{HD~30614  }&$\alpha$ Cam  &O9.5Ia      &W72&88300601    &ISO/PHe  & 85\\ 
 \object{HD~195592 }&              &O9.7Ia      &W72&90001101    &ISO/PHe  & 45\\ 
 \object{WR~147    }&              &WN8h        &S96&88000701    &ISO/PHe  & 35\\   
 \hline 	    
 \end{tabular}
 \begin{list}{}{}
 \item[$^{\mathrm{a}}$] 
 C71: Conti \& Alschuler 1971; 
 W72: Walborn 1972; 
 W73: Walborn 1973; 
 S96: Smith et al. 1996.
 \end{list}  
 \end{center}
\end{table*}
\begin{table*}[!ht]
 \begin{center}
 \caption{The 30 B-type stars observed during the ISO/SWS Post-Helium mission, 
 supplemented with 3 stars from the ISO/SWS nominal mission and 1 star observed 
 with CGS4/UKIRT.}
 \label{OB2}  
 \begin{tabular}{lclcccr}
 \hline	    
 \hline	    
 Star&Name&Spectral&Spect. Type             &ISO Observation&Instrument&S/N\\
     &    &Type    &Reference$^{\mathrm{a}}$&   Number      &          &   \\ 
 \hline	    
 \object{HD~202214 }&HR8119        &B0V         &M55&90300701    &ISO/PHe  & 15\\ 
 \object{HD~93030  }&$\theta$ Car  &B0Vp        &B62&25900905    &ISO/Nom  &115\\ 
 \object{HD~37128  }&$\epsilon$ Ori&B0Ia        &W90&            &UKIRT    &115\\ 
 \object{HD~198781 }&HR7993        &B0.5V       &M55&88301201    &ISO/PHe  & 10\\ 
 \object{HD~207793 }&              &B0.5III     &M55&88700901    &ISO/PHe  & 20\\ 
 \object{HD~185859 }&HR7482        &B0.5Ia      &M55&89901301    &ISO/PHe  &  6\\ 
 \object{HD~116658 }&$\alpha$ Vir  &B1V         &M55&25302001    &ISO/Nom  &165\\ 
 \object{HD~208218 }&              &B1III       &M55&88701101    &ISO/PHe  &  7\\ 
 \object{HD~190066 }&              &B1Iab       &M55&88101401    &ISO/PHe  & 15\\ 
 \object{HD~158926 }&$\lambda$ Sco &B1.5IV      &H82&49101016    &ISO/Nom  &140\\ 
 \object{HD~52089  }&$\epsilon$ Cma&B1.5II      &W90&88602001    &ISO/PHe  &130\\ 
 \object{HD~194279 }&V2118 Cyg     &B1.5Ia      &L92&88201301    &ISO/PHe  & 80\\ 
 \object{HD~193924 }&$\alpha$ Pav  &B2IV        &L75&88500501    &ISO/PHe  & 95\\ 
 \object{HD~206165 }&9 Cep         &B2Ib        &L68&88300301    &ISO/PHe  & 70\\ 
 \object{HD~198478 }&55 Cyg        &B2.5Ia      &L68&88100501    &ISO/PHe  &100\\ 
 \object{HD~160762 }&$\iota$ Her   &B3V         &J53&89900101    &ISO/PHe  & 70\\ 
 \object{HD~207330 }&$\pi^{2}$ Cyg &B3III       &M55&88701301    &ISO/PHe  & 45\\ 
 \object{HD~15371  }&$\kappa$ Eri  &B5IV        &H69&90701401    &ISO/PHe  & 35\\ 
 \object{HD~184930 }&$\iota$ Aql   &B5III       &L68&88000901    &ISO/PHe  & 45\\ 
 \object{HD~191243 }&HR7699        &B5II        &L92&88401401    &ISO/PHe  & 20\\ 
 \object{HD~58350  }&$\eta$ Cma    &B5Ia        &W90&90702301    &ISO/PHe  & 90\\ 
 \object{HIC~101364}&Cyg OB2 12    &B5Ia        &M91&90300901    &ISO/PHe  &105\\ 
 \object{HD~203245 }&HR8161        &B6V         &L68&88701401    &ISO/PHe  & 10\\ 
 \object{HD~155763 }&$\zeta$ Dra   &B6III       &L68&89900201    &ISO/PHe  & 80\\ 
 \object{HD~209952 }&$\alpha$ Gru  &B7IV        &H69&88500701    &ISO/PHe  &150\\ 
 \object{HD~183143 }&HT Sge        &B7Ia        &M55&89901501    &ISO/PHe  & 60\\ 
 \object{HD~14228  }&$\phi$ Eri    &B8V-IV      &H69&88701901    &ISO/PHe  & 75\\ 
 \object{HD~207971 }&$\gamma$ Gru  &B8III       &H82&88500901    &ISO/PHe  &100\\ 
 \object{HD~208501 }&13 Cep        &B8Ib        &L92&88701201    &ISO/PHe  & 80\\ 
 \object{HD~199478 }&V2140 Cyg     &B8Ia        &L92&88501801    &ISO/PHe  & 60\\ 
 \object{HD~16978  }&$\epsilon$ Hyi&B9V         &H75&88401901    &ISO/PHe  & 65\\ 
 \object{HD~196867 }&$\alpha$ Del  &B9IV        &M73&88101701    &ISO/PHe  & 75\\ 
 \object{HD~176437 }&$\gamma$ Lyr  &B9III       &J53&88401501    &ISO/PHe  &110\\ 
 \object{HD~202850 }&$\sigma$ Cyg  &B9Iab       &M55&90600601    &ISO/PHe  & 65\\ 
 \hline 	    
 \end{tabular}
 \begin{list}{}{}
 \item[$^{\mathrm{a}}$] 
  J53: Johnson \& Morgan 1953.
  M55: Morgan et al. 1955; 
  B62: Buscombe 1962; 
  L68: Lesh 1968; 
  H69: Hiltner et al. 1969; 
  M73: Morgan \& Keenan 1973; 
  L75: Levato 1975; 
  H75: Houk 1975; 
  H82: Houk 1982; 
  W90: Walborn \& Fitzpatrick 1990; 
  M91: Massey \& Thompson 1991; 
  L92: Lennon et al. 1992. 
 \end{list}  
 \end{center}
\end{table*}

\begin{table*}[!ht]
 \begin{center}
 \caption{The 14 B stars with emission lines observed in the ISO/SWS 
 Post-Helium mission supplemented with 7 stars observed during the ISO/SWS 
 nominal mission.}
 \label{Be}  
 \begin{tabular}{lclcccr}
 \hline      
 \hline      
 Star&Name&Spectral&Spect Type              &Observation &Status&S/N\\
     &    &Type    &Reference$^{\mathrm{a}}$&   Number   &      &   \\ 
 \hline          			    
 \object{V1478~Cyg }&MWC 349A      &O9III[e]  &Z98&18500704    &ISO/Nom  &145\\   
 \object{HD~206773 }&MWC 376       &B0Vpe     &M55&88502101    &ISO/PHe  & 20\\   
 \object{HD~5394   }&$\gamma$ Cas  &B0.5Ve    &P93&24801102    &ISO/Nom  &150\\   
 \object{HD~212571 }&$\pi$ Aqr     &B1Ve      &L68&90601301    &ISO/PHe  & 20\\   
 \object{HD~50013  }&$\kappa$ Cma  &B1.5IVne  &H69&90702001    &ISO/PHe  & 80\\   
 \object{HD~200775 }&MWC 361       &B2V[e]    &G68&90300501    &ISO/PHe  & 50\\   
 \object{HD~45677  }&MWC 142       &B2V[e]    &Z98&71101992    &ISO/Nom  &135\\   
 \object{HD~56139  }&$\omega$ Cma  &B2IV-Ve   &H69&90702201    &ISO/PHe  & 40\\   
 \object{HD~105435 }&HR 4621       &B2IVne    &H69&07200272    &ISO/Nom  &120\\  
 \object{HD~205021 }&$\beta$ Cep   &B2IIIe    &M55&88100301    &ISO/PHe  &110\\ 
 \object{HD~187811 }&12 Vul        &B2.5Ve    &L68&90700901    &ISO/PHe  & 25\\   
 \object{HD~191610 }&28 Cyg        &B2.5Ve    &L68&89900901    &ISO/PHe  & 35\\   
 \object{HD~205637 }&$\epsilon$ Cap&B3Vpe     &H88 &90601701    &ISO/PHe  & 30\\  
 \object{HD~10144  }&$\alpha$ Eri  &B3Vpe     &H69&90000101    &ISO/PHe  &140\\   
 \object{HD~56014  }&EW Cma        &B3IIIe    &H82 &90702101    &ISO/PHe  & 20\\   
 \object{HD~50123  }&HZ Cma        &B6Vnpe    &S   &88601901    &ISO/PHe  & 75\\  
 \object{HD~198183 }&$\lambda$ Cyg &B6IVe     &L68&89900801    &ISO/PHe  & 35\\   
 \object{HD~209409 }&omi Aqr       &B7IVe     &L68&90601501    &ISO/PHe  & 35\\   
 \object{HD~193237 }&P Cygni       &B2pe      &L68&33504020    &ISO/Nom  &100\\ 
 \object{HD~94910  }&AG Car        &B2pe      &H75&22400153    &ISO/Nom  & 80\\   
 \object{HD~93308  }&$\eta$ Car    &Bpe       &H75&07100250    &ISO/Nom  &170\\   
 \hline 	    
 \end{tabular}
 \begin{list}{}{}
 \item[$^{\mathrm{a}}$] 
M55: Morgan et al. 1955; 
L68: Lesh 1968; 
G68: Guetter 1968;
H69: Hiltner et al. 1969;  
H75: Houk 1975;
H82: Houk 1982; 
H88: Houk 1988;
P93: van Paradijs 1993;
Z98: Zorec et al. 1998;
S: Simbad.
 \end{list}  
 \end{center}
\end{table*}

 The advance of infrared-detector technology since the eighties has opened 
 new perspectives for the study of early-type stars. Investigation of the 
 early phases of their evolution especially benefits from infrared (IR) 
 observations. The birth places of massive stars are identified with 
 Ultra-Compact H{\sc ii} regions ({\sc UCHii}). In such regions, the stars are 
 still embedded in material left over from the star formation process and are 
 obscured at optical and ultraviolet wavelengths. In the K-band 
 (ranging from 2.0 to 2.4 \mum) dust optical depths $\tau$ of a few occur, 
 while in the H-band (ranging from 1.5 to 1.8 \mum) $\tau$ is typically of 
 order ten. At shorter wavelengths, the dust extinction becomes too 
  high to observe the embedded stars. The IR emission of the warm dust 
  cocoon covering the newly formed massive stars in {\sc UCHii} regions peaks 
 typically at about 100 \mum. At wavelengths longwards of 5--10 \mum, the 
 thermal emission of the dust dominates the photospheric flux, and can be as 
 much as 4 orders of magnitude above the stellar free-free 
 continuum at 100 \mum\ (Churchwell 1991). 

  Reliable values for the luminosities, temperatures and mass-loss rates of 
 the embedded massive stars are essential as they allow us to trace the very 
 early phases of their evolution of which little is known. Furthermore, 
 these parameters control the photo-dissociation and ionisation of the 
  molecular gas, the evaporation of the dust, and affect the morphology 
 of the {\sc UCHii} region.	    

  The development of {\em quantitative} diagnostics based on IR spectral 
  data requires, as a first step, homogeneous observations of a large set 
  of both normal and peculiar non-embedded early-type stars, that have been 
  studied in detail at optical and ultraviolet wavelengths where OB-type stars 
  exhibit many spectral lines. Such stars may be used to calibrate quantitative 
  methods based on IR spectroscopy alone. 
   Calibration work has already been carried out in other near-infrared
   wavelength ranges, in the J-band e.g. Wallace et al. (2000), in the H-band e.g. Meyer 
   et al. (1998) and Hanson et al. (1998), and in the K-band Hanson et al.
   (1996).
  The ``Post-Helium program'' conducted with the Short Wavelength 
  Spectrometer (SWS) on board the Infrared Space Observatory (ISO) is intended 
  to provide such a data set. 
   This mission started after helium boil-off in April 1998 and made use of the 
   ability of the detectors of SWS to acquire observations in band~1 [2.4-4.1] \mum\
   during the slow warming of the satellite (see also Sect.~\ref{ISO})
  The band~1 of ISO SWS ranges from 2.4 to 4.1 \mum, and is, 
  like the K-band, positioned favourably in the narrow window in which newly 
  born stars can be observed directly. This wavelength region contains 
  important diagnostic hydrogen lines of the Brackett (\bra, \brb), Pfund 
  (\pfg), and Humphreys series. 

  In this paper, we present and study 75 spectra of early-type stars, 
  67 [2.4-4.1 \mum] ISO/SWS spectra and 8 [3.5-4.1 \mum] spectra observed 
  with the United Kingdom Infrared Telescope (UKIRT).
  This sample includes OB, Be, and Luminous Blue Variable (LBV) stars. 
  We discuss line trends as a function of spectral type, 
  following a strategy similar to the one adopted by Hanson et al.
   (1996) for the K-band. Simple empirical methods are employed to derive 
  the spectral type and/or luminosity class. These methods may also be 
  applied if only ground-based L'-band spectra are available 
  (which cover a smaller wavelength range).

  The paper is organised as follows: In Sect.~\ref{observation} we discuss 
  the data acquisition and reduction techniques; Sect.~\ref{atlas} comprises 
  a catalogue of good quality spectra; Sect.~\ref{identification} provides 
  the line identifications. Line trends and methods to classify OB-type stars 
  are presented in Sect.~\ref{trends}, while Sect.~\ref{emmission} 
  describes the spectra of B stars with emission lines. The results are 
  summarised in the final section. The equivalent-width measurements are 
  listed in the Appendix.  

\section{Observations}
\label{observation}

 \subsection{The ISO/SWS sample}
 \label{ISO}

    The ISO spectra were obtained with ISO/SWS (SWS, de Graauw et al. 1996 ; ISO, Kessler et al. 
    1996). After helium boil-off of the ISO satellite on 8 April 1998, the near-infrared band~1 
    [2.4-4.1 \mum] of SWS equipped with InSb detectors could still be operated as the 
    temperature at the focal plane increased only slowly. Between 13 April and 10 May, 
    spectra of nearly 250 bright stars were acquired for a stellar classification program. 
    Referred to as ``Post-Helium observations'', this program aims at extending the MK 
    classification scheme into the near-infrared.
    	  
    In this paper we present the subset of O- and B-type stars observed during the Post-Helium 
    phase. These observations were executed using a dedicated engineering observation mode, 
    the so-called Post-Helium observation template. All the spectra obtained during the Post-Helium
    program, including later spectral types, as well as details about the data acquisition 
    will be published in a separate publication (Vandenbussche et al. in prep).
    Along with these Post-Helium spectra, we include ten spectra of O and B stars 
    measured during the nominal mission using Astronomical Observation Template~1 speed~4 [AOT01] . 
  
    Both observation templates use the same scanning strategy,
    SWS takes a full continuous spectrum over four preset 
    overlapping sub-bands.  These are defined in Table~\ref{prop}.
     The integration time per target is fixed, therefore the S/N ratio mainly 
    depends on the brightness of the source. 
\begin{table}
\caption{The spectral resolution R and wavelength coverage of the four sub-bands, 
for a detailed technical specification see de Grauw et al. (1996).}
\label{prop}
\begin{tabular}{ccc}
preset sub-band&$R=\Delta \lambda / \lambda$&wavelength coverage(\mum)\\
\hline
band~1a&1870-2110&2.38-2.60 \\
band~1b&1470-1750&2.60-3.02 \\
band~1d&1750-2150&3.02-3.52 \\
band~1e&1290-1540&3.52-4.08 \\
\end{tabular}
\end{table}
    	  
    Combining the nominal and Post-Helium program AOT01 speed~4 observations, we collected 69 
    ISO/SWS spectra. However, two targets (HD~147165 and HD~203245) were clearly off-pointed 
    and will therefore not be discussed. We split the remaining 67 stars into two 
    subgroups: the O- and B-type stars, and the B stars with emission-line spectra. Spectra
    over 2.4-4.1 \mum\ for the majority of these stars are presented for the first time. For 
    comparison with Of supergiants, we have included the Wolf-Rayet star WR~147 (Van der Hucht et al. 
    1996) in the first subgroup. The second 
    subgroup includes 18 Be and 3 Luminous Blue Variable (LBV) stars (see 
    Humphreys \& Davidson 1994 for a review). Spectra of AG Car and P Cyg have been presented by 
    Lamers et al. (1996a,b). The 45 OB stars are listed in Tables~\ref{OB1} 
    and~\ref{OB2} together with 8 OB stars observed with UKIRT; the 21 B stars with emission 
    lines are given in Table~\ref{Be}. 
    	  
    Each table provides the HD number and stellar name; the spectral type and luminosity class; 
    the ISO/SWS observation number and a label indicating whether the observation was done during 
    the nominal or Post-Helium program, quoted by the acronym ISO/Nom and ISO/PHe, 
    respectively. The last column provides a spectrum averaged value of the signal-to-noise 
    ratio (S/N) of the observation (see Sect.~\ref{reduction}). On average the S/N is 
    relatively low for the O- and early B-type stars: only 5 out of 22 stars of spectral type 
    earlier than B2 have a S/N $\geq$ 60; for the later type stars the situation is reversed, 
    i.e. only 5 out of 23 have S/N $\leq$ 60. This tendency is explained by the lack of 
    relatively nearby bright O and early-B stars compared to later B stars.
    For the B stars with 
    emission lines, the S/N of the continuum is not that important as the emission lines are 
    very prominent in most of the spectra. 

    The 34 B stars provide a fairly dense coverage of B spectral types, but this is not the case
    with the 12 O stars. Moreover, because of the relatively low S/N 
    of our observations, we could not detect lines in any of the five O\,V stars. Lines are 
    detected, however, in supergiant O stars. We obtained L'-band UKIRT observations in order 
    to improve the coverage of O spectral types. These are discussed in 
    Sect.~\ref{UKIRT}. The subgroup of B stars with emission-line spectra shows a diversity 
    in the way their circumstellar material is distributed: 18 Be stars with discs and/or 
    shells and 3 LBV stars ($\eta$~Carinae, AG~Carinae and P~Cygni) with dense stellar winds.
  \begin{figure*}[!ht]
    \begin{center}
    \resizebox{17cm}{!}{\includegraphics{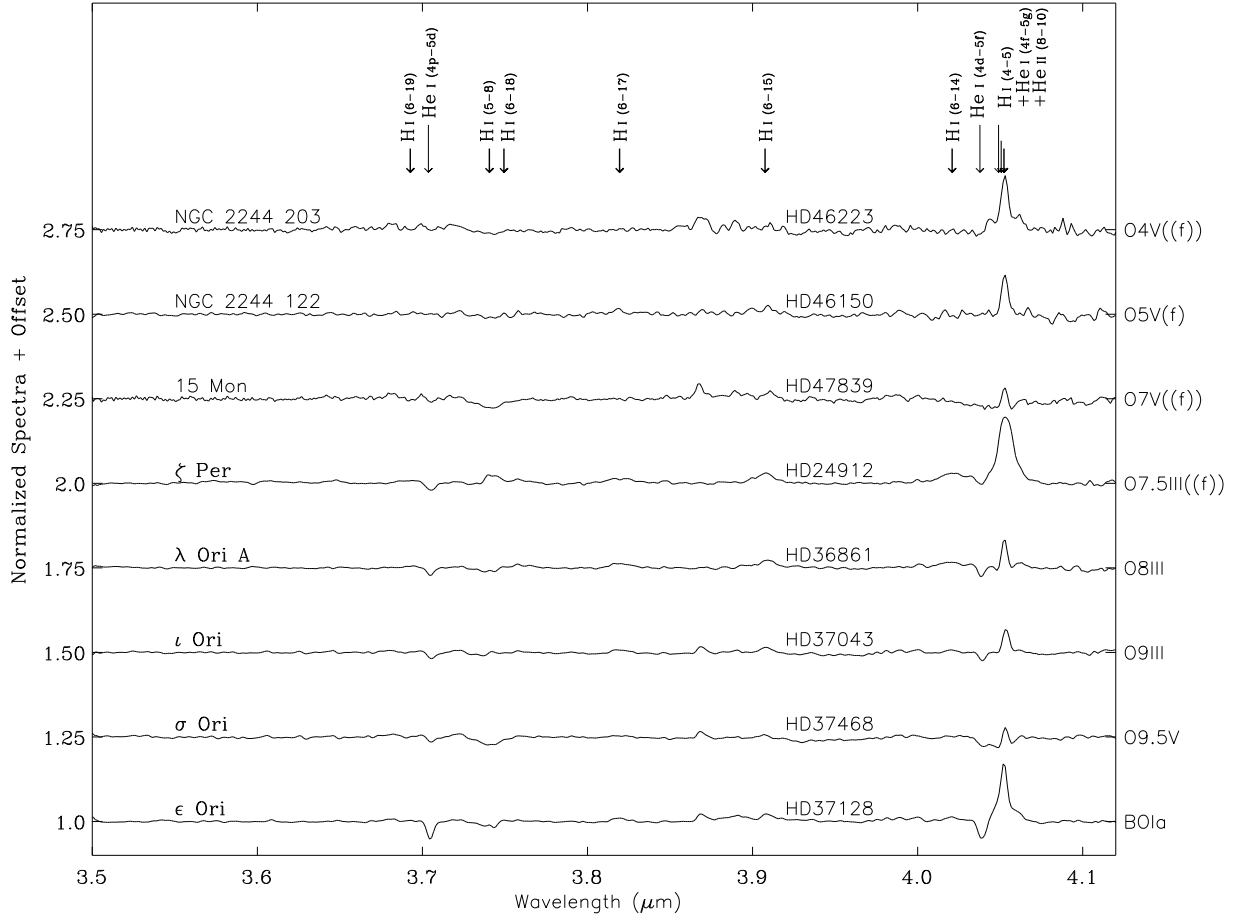}}
    \caption{The 3.5 to 4.12 \mum\ region of the spectra of O- type stars obtained with 
    CGS4/UKIRT. All O stars in this sample show \bra\ emission in the core.}
    \label{UK}    
    \end{center}  
  \end{figure*} 

 \begin{figure*}[!ht]
    \begin{center}
    \resizebox{17cm}{!}{\includegraphics{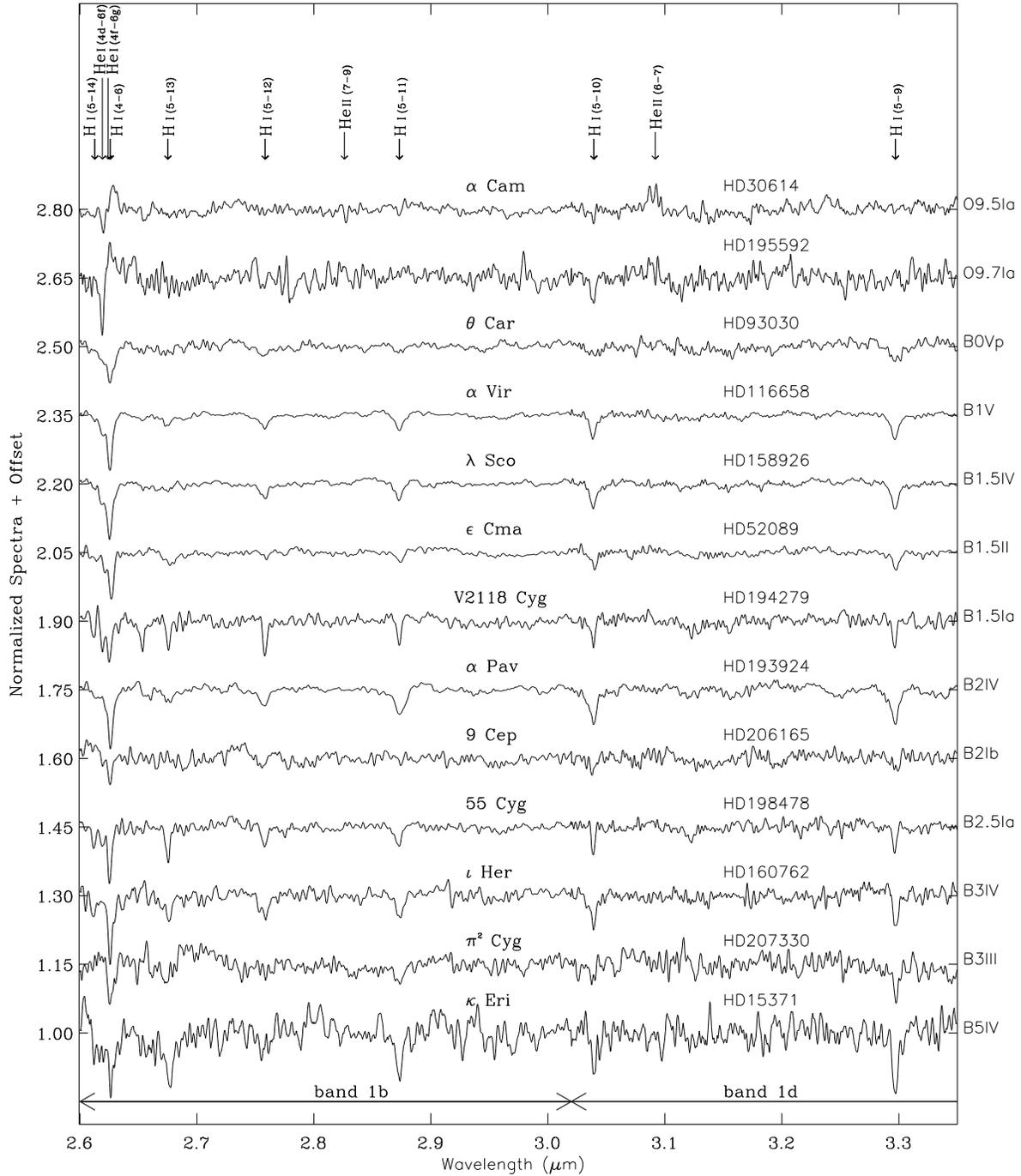}}
    \caption{The 2.6 to 3.35 \mum\ region of the spectra of O9- to B5-type stars contains the 
    \brb\ line at \lam~2.6259 \mum\ and some of the Pfund series lines. A few helium lines are 
    detected in the hottest stars and are identified with arrows in the top of the figure. The O 
    supergiants show \brb\ emission.}
    \label{spectrumOB1}
    \end{center}  
  \end{figure*} 
		  
  \begin{figure*}[!ht]
    \begin{center}
    \resizebox{17cm}{!}{\includegraphics{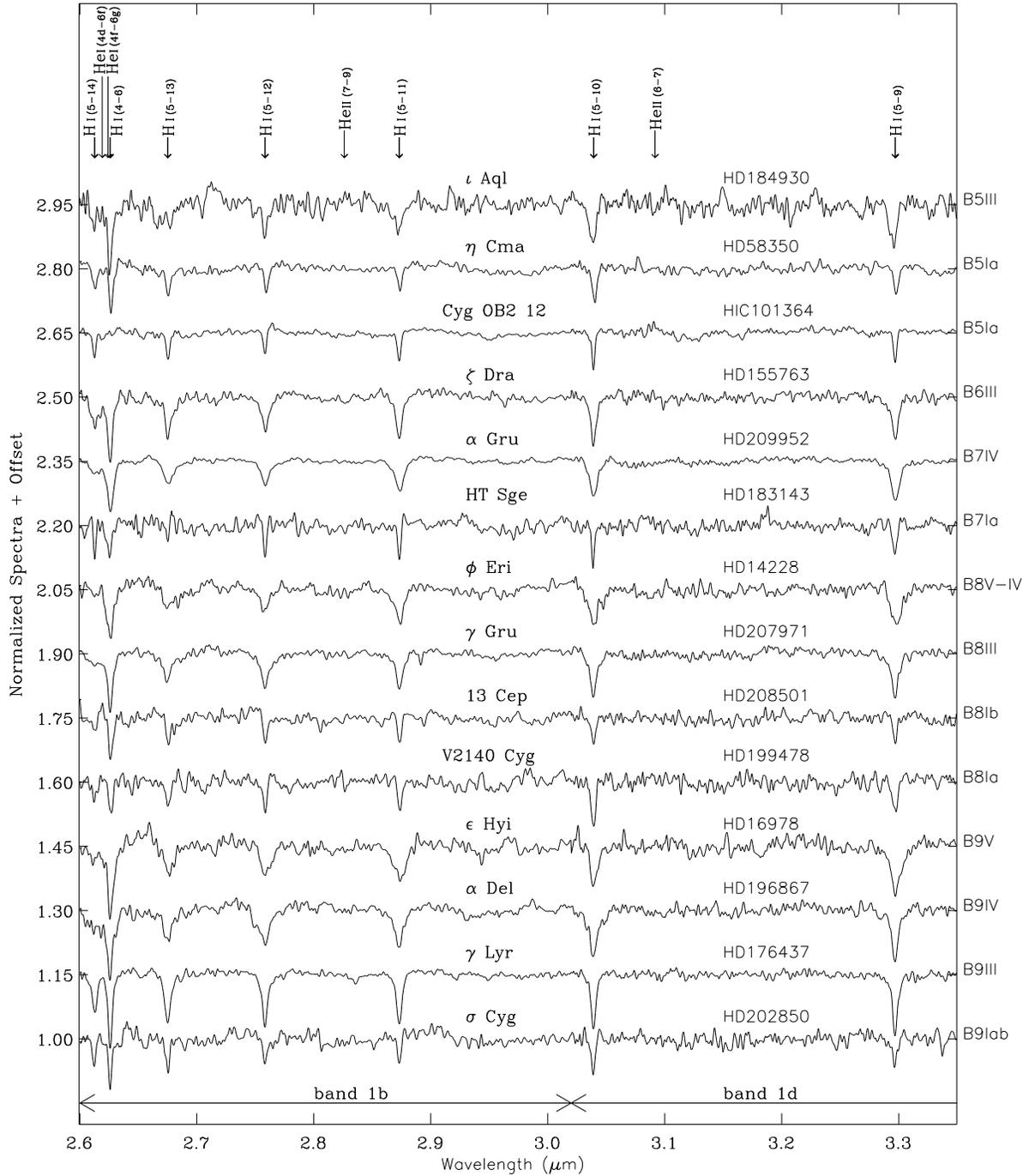}}
    \caption{The 2.6 to 3.35 \mum\ region of the spectra of B5- to B9-type stars contains the 
    \brb\ line at \lam~2.6259 \mum\ and some lines of the Pfund series. Helium lines are no longer 
    present.}
    \label{spectrumOB2} 
    \end{center}  
  \end{figure*} 
		  
  \begin{figure*}[!ht]
    \begin{center}
    \resizebox{17cm}{!}{\includegraphics{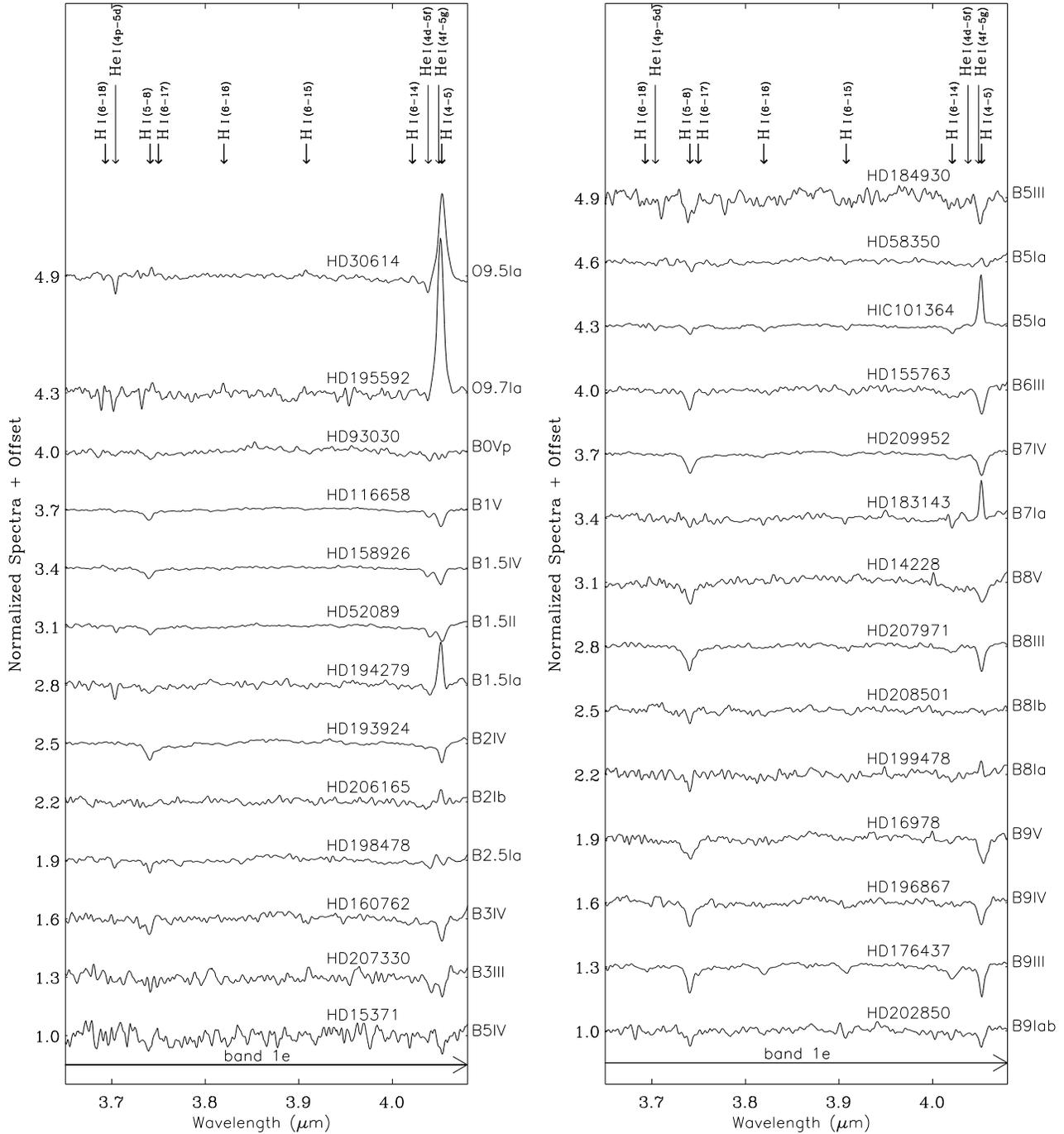}}
    \caption{The 3.65 to 4.08 \mum\ region of the spectra of O9- to B9-type stars contains the 
    \bra\ line at \lam~4.0523 \mum\ and some of the Humphreys series lines. \hei\ lines are 
    present down to spectral type B3 and are identified with arrows in the top of the figure. 
     The OB supergiants show \bra\ emission down to B7/8.}
    \label{spectrumOB3}
    \end{center}  
  \end{figure*}

  \begin{figure*}[!ht]
    \begin{center}
    \resizebox{17cm}{!}{\includegraphics{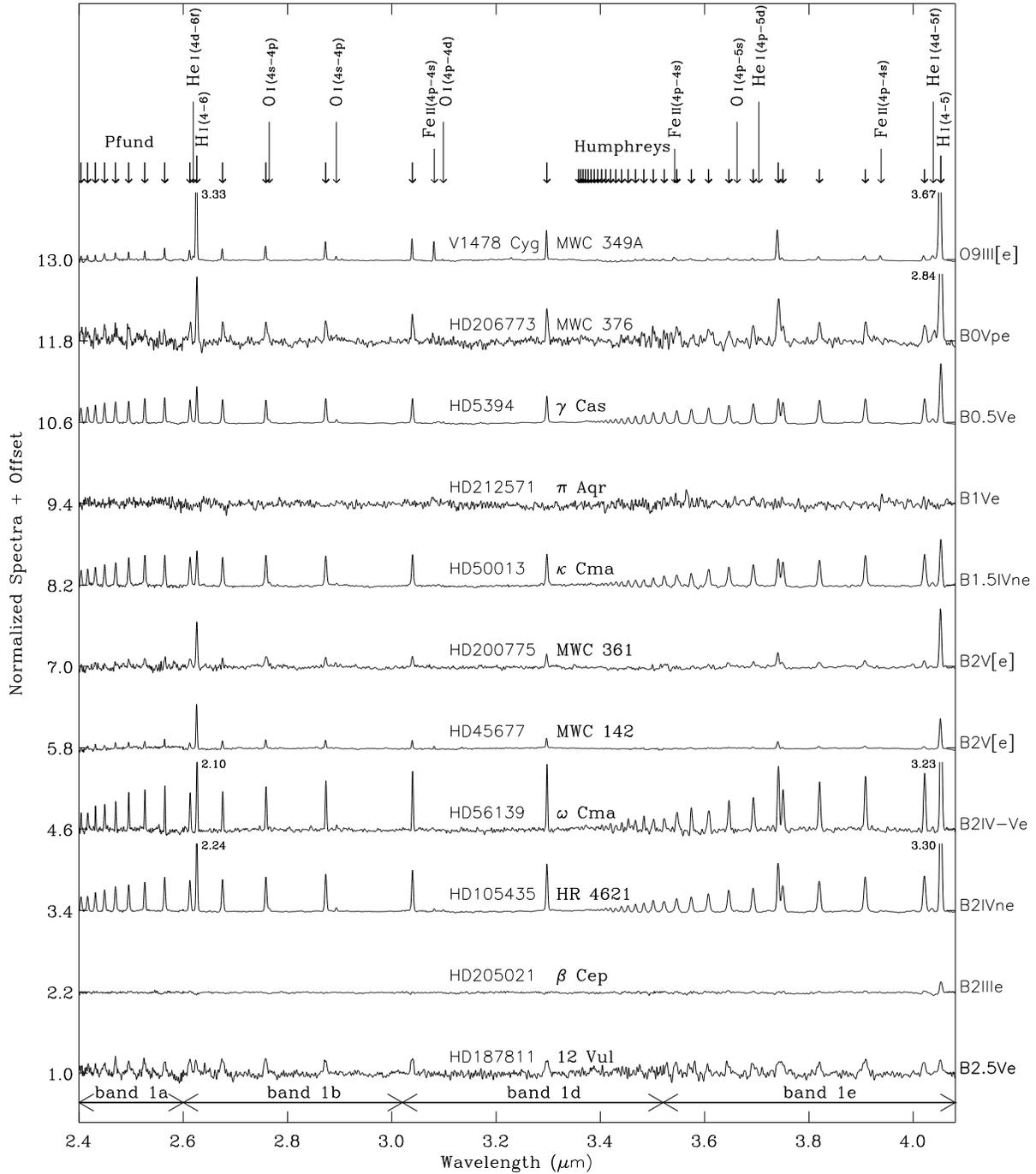}}
    \caption{The full band~1 spectra from 2.40 to 4.08 \mum\ of B0e to B2.5e stars contain 
    hydrogen lines of the Brackett, Pfund and Humphreys series. In most of the spectra, the 
    only \hei\ line present is at \lam4.0377. A few stars also show some \oi\ lines.}
    \label{Be1}   
    \end{center}  
  \end{figure*} 
    		      
  \begin{figure*}[!ht]
    \begin{center}
    \resizebox{17cm}{!}{\includegraphics{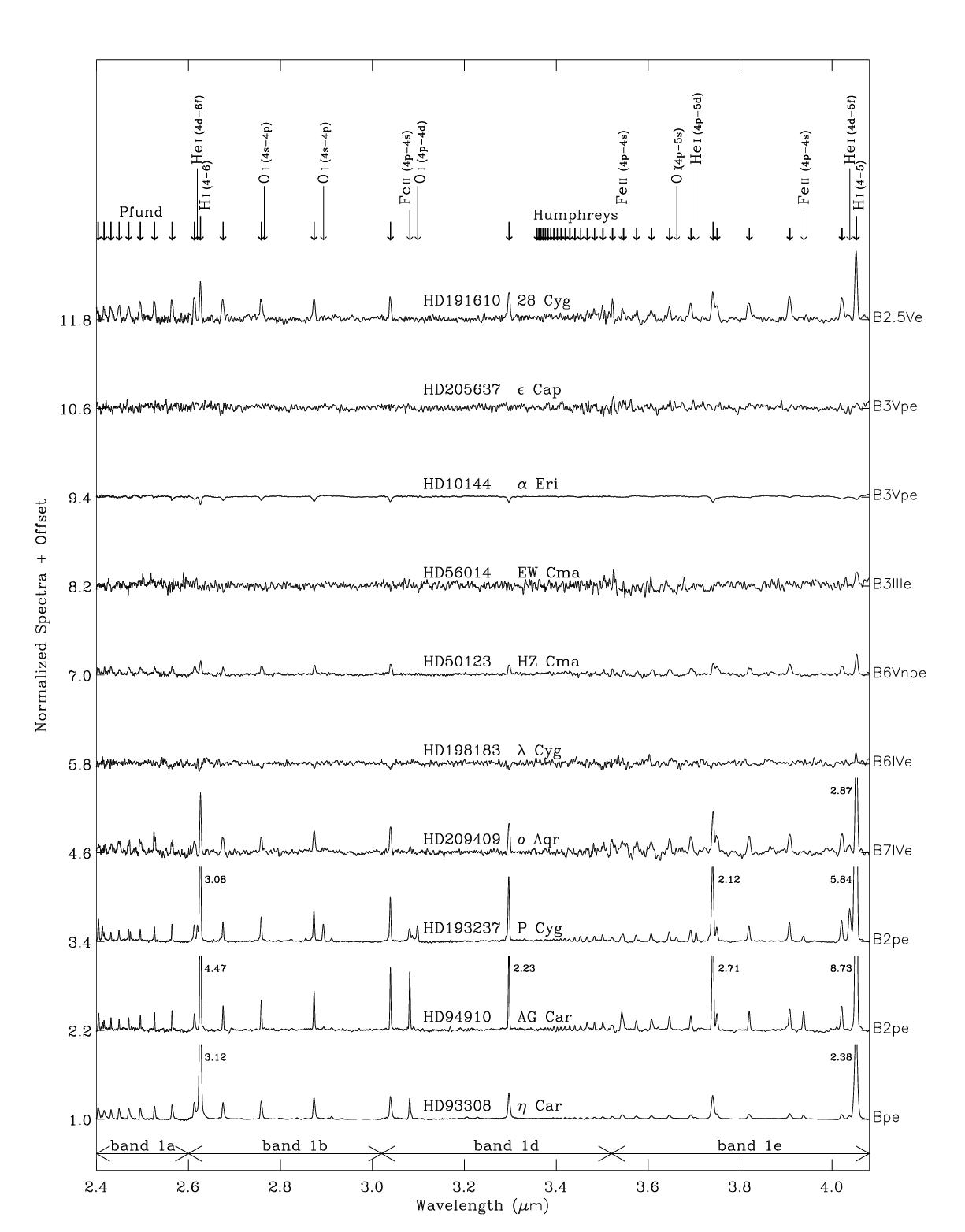}}
    \caption{The full band~1 spectra from 2.40 to 4.08 \mum\ of B3e to B7e stars contain only 
    hydrogen lines. HD~10144 does not currently show emission lines in its spectrum. The three 
    Luminous Blue Variables, as well as the three peculiar Bpe stars, show some lines of \hei, 
     \feii, \mgii\ and \oi.}
    \label{Be2}   
    \end{center}  
  \end{figure*} 
  \subsubsection{ISO/SWS data reduction}
  \label{reduction}

    The data acquired during the nominal mission were calibrated in the SWS Interactive Analysis 
    environment with the calibration files as in Off-Line Processing Version 10.0. The 
    Post-Helium data required special care as changes in the characteristics of the instrument
    arose when the temperature increased. A time-dependent calibration was derived, based on 
    reference observations in each orbital revolution of the satellite. This accounts for 
    changes in wavelength calibration and photometric sensitivity as a function of wavelength. 
    Fortunately, the spectral resolution did not change and the dark current and noise remained 
    fairly similar, 
    as the signal registered with closed instrument shutter is still dominated by the amplifier
    offsets. The exact sources of instrumental drifts cannot be fully disentangled but a 
    reliable empirical calibration could be derived. The Post-Helium calibration, which is 
    described in detail in Vandenbussche et al. (2000), results in a data quality that is 
    comparable to that during the nominal mission. 
    To illustrate this: P~Cyg was observed both during the nominal and Post-helium missions, 
    the spectra show a continuum level variation of 4 \% and a line width variation of 5 \%. 
 
    All the spectra were processed from the Auto-Analysis Result stage using the SWS 
    Interactive Analysis ($IA^{3}$) programs. First, the behaviour of the individual detectors 
    was checked. Second, the two independent spectral scans were compared. 
    Discrepancies were treated when their cause was clearly established 
    (jumps, glitches, residual tilt in the slope of the Post-Helium spectra). 
     The adopted spectral resolution per sub-band is very similar to the R values given in 
    Sect.~\ref{ISO}, but not strictly identical, as the final rebinning is based on on-board
    measurements (see Lorente et al. 1998 and Hony et al. 2000).
  \subsection{The CGS4/UKIRT sample}
  \label{UKIRT}

   The UKIRT spectra were obtained on the second half of the night of 23
   December 2000 (UT) using the Cooled Grating Spectrometer 4 (CGS4; Mountain
   et al. 1990). We obtained L'-band (3.5-4.1\,\mum) spectra of 8 stars with
   spectral types ranging from O4 to B0. The 40 l/mm grating was used in
   first order with the 300\,mm focal length camera and the 0.6 \arcsec wide slit,
   giving a nominal resolution of 0.0025\,\mum\ (R$\approx$1500). The array was stepped to
   provide 2 data points per resolution element. Signal-to-noise ratios of 70
   to 200 were achieved on the continua of the target hot stars. The four O V
   stars, with subtypes O4, O5, O7, and O9.5 significantly improve the
   coverage of spectral types. Three O7 to O9 giants were also observed, as
   well as one B0 supergiant.

   For data reduction, we used the Starlink Figaro package. Spectra were
   ratioed by those of dwarf A and F stars observed on the same night at
   similar airmasses as the hot stars, corrected for the approximate
   effective temperatures of the stars by multiplying by a blackbody
   function. Wavelength calibration was achieved using the second order
   spectrum of an argon arc lamp. The spectra shown here have been slightly
   smoothed, and have a resolution of 0.0031\,\mum\ (R$\approx$1200).
  \subsection{Atlas}
  \label{atlas}

    We present the normalised ISO/SWS spectra of O and B stars with S/N greater than 30 from 
    2.6 to 3.35 \mum\ and from 3.65 to 4.08 \mum\ in Fig.~\ref{spectrumOB1} to 
    \ref{spectrumOB3}. We do not display the band 1a (from 2.4 to 2.6 \mum) because
    the S/N of this sub-band, containing the higher Pfund series and for two stars only
    a probable \siiv\ line, is significantly lower than for the others.
    The spectra from 3.35 to 3.65 \mum\ do not show any detectable lines.
    Fig.~\ref{UK} displays the L'-band spectra obtained with CGS4/UKIRT. 
    Fig.~\ref{Be1} and \ref{Be2} display the full ISO/SWS band~1 spectra of all Be and 
    Luminous Blue Variables stars in our sample. Line identifications are provided in each 
    of the figures.

\section{Identification and measurement of spectral lines}
\label{identification}

  \begin{table*}[!ht]
   \caption{Lines identified in the [2.4-4.1] \mum\ region.}
    \label{linelist}
    \begin{tabular}{llc|llc|llc}
    \hline	     
     $\lambda_{\rm vac}$&Element&Configuration&$\lambda_{\rm vac}$&Element&Configuration
     &$\lambda_{\rm vac}$&Element&Configuration \\
    (\mum)&&&(\mum)&&&(\mum)&&\\   
    \hline
      2.404&\hi&$5-22$&                 2.826&\heii&$7-9$&                 3.546&\hi&$6-22$\\                    
      2.405&\mgii&$4d-5p$&	         2.873&\hi&$5-11$&                  3.574&\hi&$6-21$\\                   
      2.413&\mgii&$4d-5p$&	         2.893&\oi&$2s^22p^34s-2s^22p^34p$& 3.607&\hi&$6-20$\\                   
      2.416&\hi&$5-21$&		 3.039&\hi&$5-10$&                  3.646&\hi&$6-19$\\                   
      2.427&\siiv&$4d-4f$&	         3.081&\feii&$3d^64p-3d^64s$&       3.662&\oi&$2s^22p^34p-2s^22p^35s$\\  
      2.431&\hi&$5-20$&		 3.092&\heii&$6-7$&                 3.693&\hi&$6-18$\\                   
      2.449&\hi&$5-19$&		 3.095&\heii&$8-11$&                3.704&\hei&$1s4p-1s5d$\\             
      2.450&\hi&$5-18$&		 3.098&\oi&$2s^22p^34p-2s^22p^34d$& 3.741&\hi&$5-8 $\\                   
      2.473&\hei&$1s4p-1s6d$&	         3.297&\hi&$5-9 $&                  3.749&\hi&$6-17$\\                   
      2.495&\hi&$5-17$&		 3.402&\hi&$6-32$&                  3.819&\hi&$6-16$\\                   
      2.526&\hi&$5-16$&	      	 3.410&\hi&$6-31$&                  3.907&\hi&$6-15$\\                   
      2.564&\hi&$5-15$&		 3.419&\hi&$6-30$&                  3.938&\feii&$3d^64p-3d^64s$\\        
      2.613&\hi&$5-14$&		 3.429&\hi&$6-29$&                  4.021&\hi&$6-14$\\                   
      2.620&\hei&$1s4d-1s6f$&	         3.440&\hi&$6-28$&                  4.038&\hei&$1s4d-1s5f$\\             
      2.624&\hei&$1s4f-1s6g$&  	 3.453&\hi&$6-27$&                  4.041&\hei&$1s4d-1s5f$\\             
      2.625&\heii&$8-12$&	         3.467&\hi&$6-26$&                  4.049&\hei&$1s4f-1s5g$\\             
      2.626&\hi&$4-6 $&		 3.483&\hi&$6-25$&                  4.051&\heii&$8-10$\\                 
      2.675&\hi&$5-13$&		 3.501&\hi&$6-24$&                  4.052&\hi&$4-5 $ \\                      
      2.758&\hi&$5-12$&		 3.522&\hi&$6-23$&                        \\
      2.765&\oi&$2s^22p^34s-2s^22p^34p$&3.542&\feii&$3d^64p-3d^64s$&\\      				       
    \hline	     
    \end{tabular}
  \end{table*}

  In this section, we give an overview of the lines observed in the 2.4 to 4.1 \micron\ 
  region and review how we measured line strengths and widths. The investigated spectral range 
  is dominated by lines of hydrogen and helium. We made a special effort to identify lines of 
  other elements, resulting in the detection of only one silicon emission line in two late 
  O supergiants and a few lines of oxygen, magnesium and iron, in the sample of B stars with 
  emission lines. 
  \subsection{Overview of lines in the 2.4 to 4.1 \mum\ region}
  \label{lines}

    Hydrogen lines of three different series are present in this wavelength
    region: the two leading lines of the Brackett series \bra~\lam4.0523 
    (wavelength in \mum\ ) and \brb~\lam2.6259; the Pfund series 
    from \pfg~\lam3.7406 to Pf(22-5)~\lam2.4036, and the higher members of the 
    Humphreys series starting from transition \hu~\lam4.0209. The lower members 
    of each series, such as \bra, \brb\ and \pfg, are expected to be particularly 
    important diagnostic lines.
    			      
    Lines of ionised helium are identified in three O supergiant stars. 
    The (7-6) transition at \lam3.0917 is expected to be the strongest \heii\ 
    line in the H, K and L'-bands. A second strong \heii\ line, 
    (9-7) at \lam2.8260, is detected in the spectrum of the early-O 
    supergiant \object{HD~190429} and possibly in \object{HD~188001} and 
    \object{HD~30614}. It is likely that \heii~(10-8), (11-8) and (12-8) are present
    in the spectrum of HD~190429 based on a comparison with \object{WR~147}, 
    but as these lines are located in the wings of the much stronger 
    \bra, \heii~(7-6) and \brb\ lines, respectively, 
    we cannot provide a positive identification. 
    			      
    The neutral helium line that is expected to be the strongest is \hei~(5d-4f)
    at \lam4.0490. Unfortunately, this line is blended with \bra. The second 
    strongest \hei\ line in band~1 is the (5f-4d) transition at 
    \lam4.0377. This line is observed in absorption in stars from spectral type O9.5 down to 
    B2.5 and in emission in Be stars of similar 
    spectral type. Of comparable strength are \hei~(5d-4p)~\lam3.7036 and 
    (6f-4d)~\lam2.6192. One would also expect, \hei~(6g-4f)~\lam2.6241, but
    this line is blended with \brb\ and could not be detected.
    			      
    We found an emission line at \lam2.4275 in the two good 
    quality spectra of the late-O supergiants HD~30614 and HD~195592, 
    the most likely identification being \siiv (4f-4d).
    A few permitted \oi\ as well as \feii\  and likely \mgii\ lines could be 
    identified in several Be stars and/or LBVs. 
    \feii~(4s-4p) at \lam3.0813, \lam3.5423 and \lam3.9378 is present in all three 
    LBVs as well as in a few Be stars. \mgii~(5p-4p) at \lam2.4048 and 
    \lam2.4131 is possibly identified in all three LBVs. These identifications
    are consistent with the K-band spectra for the same stars,
    see Hanson et al. (1996).  
    Finally, four neutral oxygen lines are seen in early Be stars as well as in two LBVs: 
    \oi~(4p-4s) at \lam2.764 and \lam2.893, \oi~(5s-4p) at \lam3.662 and 
    \oi~(4d-4p) at \lam3.098.
    All identified lines are listed in Table~\ref{linelist}.

    A few forbidden lines are also observed in the spectra of LBV's and WR. We 
    did not investigate those lines here, a listing of those can be found 
    in Lamers et al. (1996b) and Morris et al. (2000). 
  \subsection{Of supergiants and \object{WR~147}}
  \label{Of}

  \begin{figure*}[!ht]
    \begin{center}
    \resizebox{17.5cm}{!}{\includegraphics{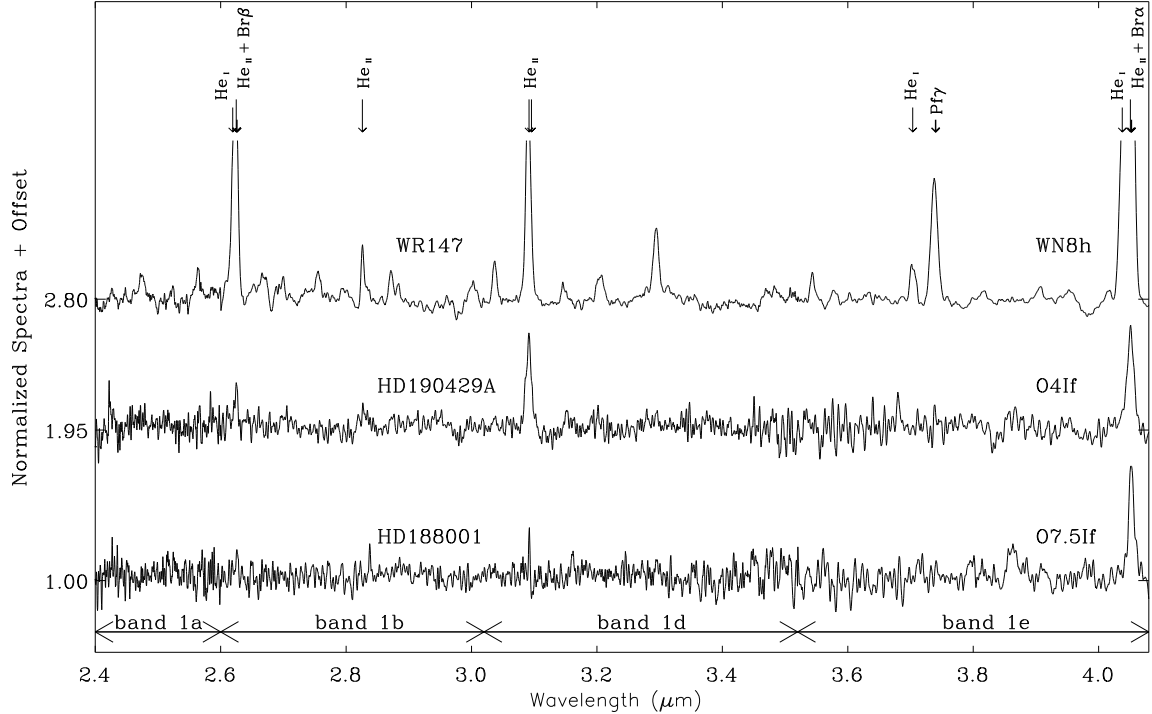}}
    \caption{Comparison of \hi\ and \heii\ lines between early Of-type 
    supergiants and WR~147 in the 2.4 to 4.1 \mum\ region. Line ratios such as \brb/\bra\ 
    and \pfg/\bra\ are roughly similar; however \heii\,(7-6)\,\lam3.092/\bra\ in HD~190429 
    is stronger by a factor of three compared to WR~147, being consistent with the higher 
    temperature of the O4If star (see Conti\& Underhill 1988).}
    \label{spectrumWR}
    \end{center}  
  \end{figure*} 

    The spectra of the two Of supergiants in our sample are plotted in Fig.~\ref{spectrumWR} 
    together with the spectrum of the Wolf-Rayet star WR~147. The ISO/SWS spectrum of WR147 has been 
    analysed in detail by Morris et al. (2000). The line strengths in the Of 
    spectra are significantly less than in the spectrum of WR~147, which is mainly a result 
    of the higher density of the wind of the Wolf-Rayet star. Line ratios such as \brb/\bra\
    and \pfa/\bra\ are roughly similar for both the Of stars and the WN8h, indicating the 
    primary dependence of the line on mass flux $\mdot/4\pi\rstar^{2}$. However, 
    the \heii~(7-6)/\bra\ line in HD~190429 is stronger by a factor of three 
    compared to WR~147, indicating that this O4 star is 
    significantly hotter. The higher temperature of the O4f stars is also implied by the 
    absence of He~{\sc i} lines. A distinction between these types seems possible on the basis 
    of overall line strength of the spectra (cf. Morris et al. 1997), though further investigation 
    of WR spectral characteristics in the near-infrared is still needed to more firmly establish 
    Of/WN differences (as in the K-band study of ``transition'' spectra by Morris et al. 1996) 
    and connections.
    		
    Concerning the O7.5\,If star, the Brackett lines are weaker and narrower than in the O4\,If 
    star indicating a lower mass-loss rate. Again, we do not detect \hei\ lines; the 
    narrow feature at the position of the \heii\ lines might be spurious. All other features between 3.4 
    and 4.0 \mum\ are due to noise.  
    \begin{figure*}[!ht]
    \resizebox{17.5cm}{!}{\includegraphics{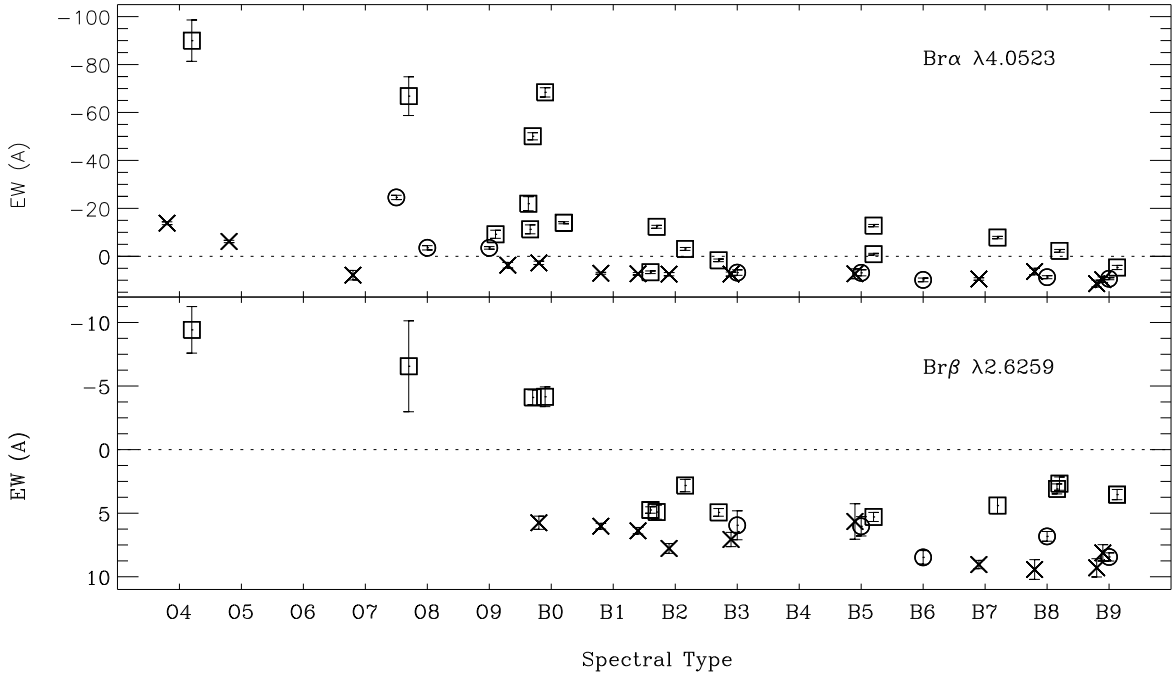}}
    \caption{The equivalent widths of \bra\ (top panel) and \brb\ (bottom panel) for normal 
    B-type stars. Stars of luminosity classes Ia-II are denoted by square symbols; class III 
    by circles, and classes IV-V by crosses. The dotted lines indicate where the lines revert 
    from absorption ($\weq > 0$) to emission ($\weq < 0$). For both lines, the B dwarfs and 
    sub-giants show a gradual increase in absorption strength towards later spectral type. 
     This can not be seen clearly in this figure, however the linear fit parameters of
    this trend are given in Table~\ref{dwarfs}. In B supergiants and bright giants the line strength remains about constant, albeit 
    with a large scatter.}
    \label{brackett}
    \end{figure*}
    		        		    
    \begin{figure*}[!ht]
    \resizebox{17.5cm}{!}{\includegraphics{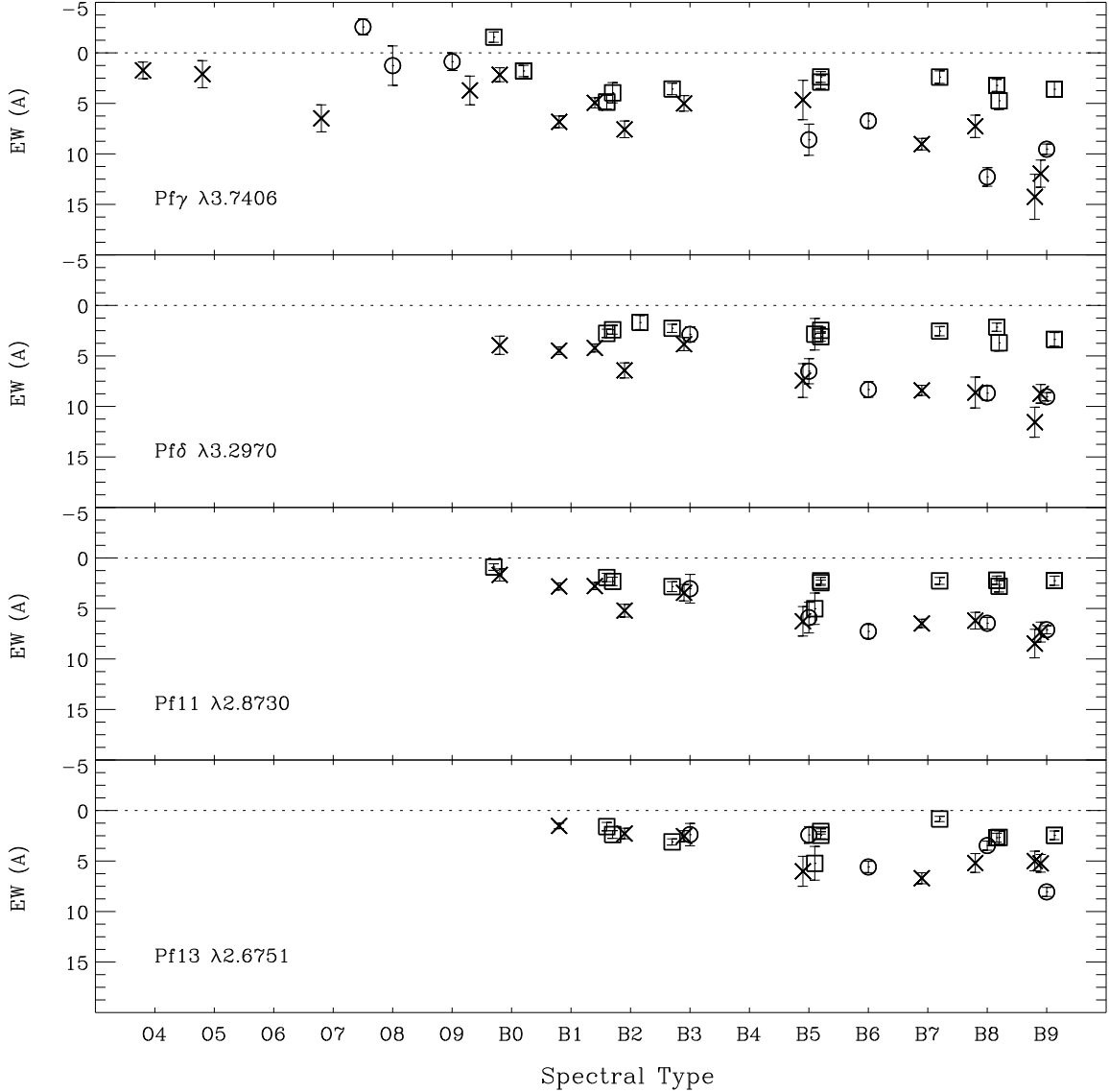}}
    \caption{The equivalent widths of four Pfund series lines for normal O and B-type 
    stars. The symbols have identical meaning as in Fig.~\ref{brackett}. Like the Bracket 
    series lines, these Pfund lines show a different behaviour for B stars of luminosity
    classes Ia-II compared to classes IV-V, i.e. the latter show a gradual increase in 
    absorption strength towards later spectral type, while in the former the strength remains 
    about constant.}
    \label{pfund}
    \end{figure*}

  \subsection{Equivalent width}
  \label{EW}

    For consistency in the measurements of equivalent widths, we first rebin the UKIRT spectra 
    to the resolution of ISO/SWS. We then define the continuum regions after removing all the 
    spectral sections containing identifiable lines. A normalisation function of the form 
    $A_{0}+A_{1} \times X +A_{2} \times X^{A_{3}}$ is fitted to each of the 4 sub-bands.
    The S/N is computed as being the inverse of the standard deviation on the normalised 
    continuum. The line parameters, equivalent width (EW) and full width at half maximum (FWHM)
    are measured on the normalised spectra using the ISO Spectral Analysis Package. 
    The errors on those measurements are dominated by the uncertainty in the position 
    of the continuum,  which is $\sim 5$ \% (Decin et al. 2000). For unblended lines, 
    the tool {\sc moment} is used as it gives 
    statistical parameters without making any assumptions on the shape of the profile. 
    		
    The signal-to-noise ratio and the spectral resolution of the ISO/SWS sample may vary over 
    the spectrum (up to 50 \%), as well as within a sub-band. This is largely intrinsic to the 
    instrument setting and depends little on the difference in flux over wavelengths; the S/N 
    varies inversely to the spectral resolution.
    		
    The EWs of the lines used in our analysis (see Sect.~\ref{trends}) are presented in 
    the Appendix. Concerning the O and B stars, this includes lines from all spectra that have 
    S/N $\ge 35$. For a few bright giant and supergiant O-stars with signal-to-noise 
    ratios smaller than this value, line measurements are presented for the relatively strong 
    \bra\ profile, and in the case of HD~190429 (O4I) and QZ~Sge (O7.5Ia) for the \brb\ and 
    \heii~(6-7) and (7-9) transitions. For one B star, HD~191243 (B5II), only three Pfund series 
    lines could be measured. This is due to a poorer S/N of sub-band 1a compared to 
    sub-bands 1b and 1d.
    		
    In the Be and LBV subgroups, lines could be measured with sufficient accuracy in all but 
    two stars, $\epsilon$~Cap (B2.5Vpe) and $\pi$~Aqr (B1Ve). The S/N of those two observations
    is quite low, 30 and 20, respectively, and the lines are not sufficiently prominent to be 
    detectable in our ISO/SWS spectra.

    \begin{figure}[!ht]
    \resizebox{8.5cm}{!}{\includegraphics{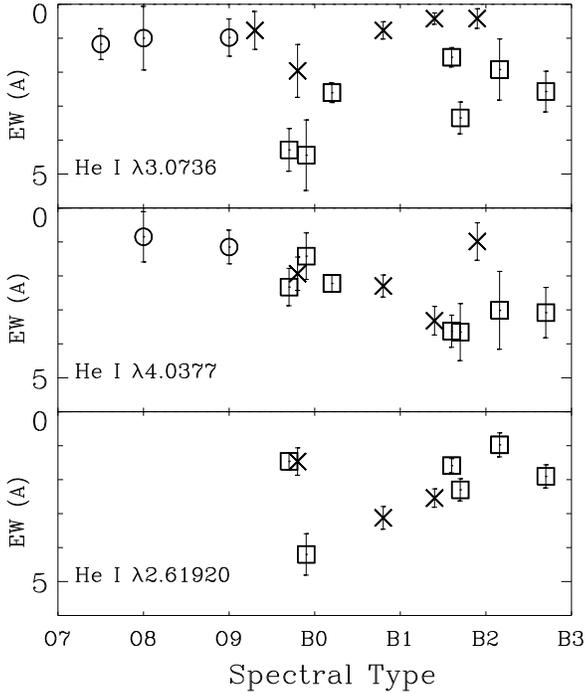}}
    \caption{The equivalent width of three \hei\ lines for normal O and B-type stars. The 
    symbols have identical meaning as in Fig.~\ref{brackett}. The \lam~3.7036 shows a 
    luminosity-class dependence.}
    \label{helium}
    \end{figure}
\section{Line trends and spectral classification of O and B stars}
\label{trends}

  In discussing the trends in line strengths of O and B stars we separately consider luminosity classes
  Ia-II and III-V, because the behaviour of the hydrogen lines in the two groups is different. 
  The difference is almost certainly connected to 
  the density of the stellar wind. In main-sequence stars, which have weak winds, the 
  line strength is dominated by temperature effects. As in optical spectra, one expects 
  a gradual weakening of the lines towards higher effective temperatures. In supergiant 
  stars, which have dense winds, the strength of the lines connecting lower levels 
  of a series (such as \bra) are expected to be highly sensitive to the stellar mass-loss rate 
  \mdot, or better stated, to the stellar mass flux $\mdot/4\pi\rstar^{2}$. Indeed, in our data 
  set \bra\ reverts from a strong absorption profile in B giants and dwarfs to a strong 
  emission profile in B supergiants, suggesting that the line is sensitive to mass loss.
  The equivalent widths of the hydrogen lines are presented in Fig.~\ref{brackett} 
  for the Brackett lines and in Fig.~\ref{pfund} for the Pfund lines. In these figures, 
  the luminosity classes Ia-II are denoted by a square (and plotted slightly to the right 
  of their spectral type); class III by a circle, and classes IV-V by a cross (and 
  plotted slightly to the left of their spectral type). In order to quantify the behaviour of 
  hydrogen lines with spectral type, we assign values to spectral types. Spectral types B0 to 
  B9 are assigned the values 10 to 19. 
  For the B-type dwarfs to giants, a quantitative trend is then derived by fitting the 
  EW versus spectral type
  with a first-order polynomial of the form $A \times S.T. + B$, where S.T. ranges 
  between 10 and 19 as defined above. This is done for 6 
  dominant hydrogen lines: \bra, \brb, \pfg, \pfd, \pff\ and \pfh, the results 
  are presented in Table~\ref{dwarfs}.	We did not measure 
  Pf(10-5), nor Pf(12-5) as we decided to focus on the behaviour among a wide range of 
  upper levels. Hydrogen lines from higher transitions are too weak to be measured in 
  a significant fraction of our sample. We do not extend the same strategy to O-type 
  stars. Indeed, the extrapolation of the linear trend we apply for B-type stars does 
  not provide a satisfactory fit to the data points for O-type stars. The O-type 
  sample is too small to build a quantitative scheme of spectral classification. 
  Moreover, at least one O stars of our sample cannot be part of a general analysis of 
  characteristics of normal early-type stars. Indeed,  the O7\,V star HD~47839 is a 
  spectroscopic binary. The early B companion affects the spectrum significantly, 
  making the hydrogen lines broader and stronger (Gies et al. 1997). 
  Therefore O-type stars are discussed in a more qualitative way in Sect.~\ref{Ostars}.
  \subsection{B supergiants and bright giants}
  \label{Bsupergiants}


    \bra\ is mostly in emission while \brb\ is the strongest in absorption of all the hydrogen lines 
    observed, the others getting weaker with higher series and members.			
    The hydrogen lines of B supergiants do not show a significant spectral-type dependence 
    but remain roughly constant, although with a large scatter (see Fig.~\ref{brackett} and 
    \ref{pfund}). This may be related to the variable nature of relatively strong lines in B 
    supergiants. Outward propagating density enhancements (spectroscopically identified as 
    discrete absorption components) and/or modulation of the overall mass-loss rate has been 
    suggested as causes for the time variability of line strength and line shape (see Kaper 
    1998 for a review). For instance, Kaufer et al. (1996) suggest, on the basis of time-series 
    analysis of H${\alpha}$ in B- and A-type supergiants, that observed variations are due to 
    rotational modulation possibly induced by weak magnetic surface structures, stellar pulsations,  
    and/or instabilities of the ionisation structure of the wind. In dwarf stars, the profiles are 
    predominantly formed in the photosphere where these phenomena are expected to have only a 
    minor impact on the line strength. Therefore, in dwarfs a dependence of line 
    strength on spectral type may be expected (see Sect.~\ref{Bdwarfs}).
    	
     Neutral helium lines are detected in O9.5-B3 stars, and therefore, can be used to 
    constrain the spectral type to earlier than B3. In the two O supergiants in our sample, 
    the S/N is unfortunately too poor to detect \hei. We did not attempt to use the line 
    strength to set the sub-type within O9.5-B3 to avoid over-interpretation.		
    We note that the \hei\ line \lam3.0736 \mum\ is found to be systematically stronger 
    in supergiants than in dwarfs stars (cf. Fig.~\ref{helium}). 
  \subsection{B dwarfs and giants}
  \label{Bdwarfs}

    \begin{table}[!ht]			   
    \caption{Fits giving the relation between spectral type S.T. and
       EW (in \AA) of Brackett and Pfund lines in B-type dwarfs
       to giants. For zero and first-order polynomials, we give
       the fit coefficients A (in \AA/S.T.) and B (in \AA) 
       and their errors, as well as 
       $\sqrt{\chi^{2}}/N$ as a measure for the goodness of fit
       (which should be less than about unity). N is the number of stars for 
       which data is available. `All' denotes the sum of all the
       individual lines given in the table; `Pfund' for the sum
       of Pf$\gamma$, Pf$\delta$, Pf(11-5), and Pf(13-5), and
       `L'-band' refers to the sum of \bra\ and \pfg. The most accurate
       spectral types may be derived from the `Pfund' lines.} 
    \begin{tabular}{lcccccc}	   
    \hline	   			   
    line      &A&dA&B&dB&$\sqrt{\chi^{2}}/N$&N\\   
    \hline	   			   
    \bra      &--   &  --  &7.93 &0.18&0.66&15\\ 
              &0.41&0.05&1.98 &0.79&0.41&15\\ 
    \hline
    \brb      & --  & --   &7.20 &0.11&0.74&15\\
              &0.29&0.03&3.11 &0.49&0.46&15\\
    \hline
    \pfg      & -- &--    &7.21 &0.20&0.93&14\\
              &0.65&0.06&-2.23&0.91&0.53&14\\
    \hline
    \pfd      & --   &  --  &6.29 &0.16&0.93&15\\
              &0.65&0.05&-3.03&0.74&0.36&15\\
    \hline
    \pff      &  --  & --   &4.78 &0.15&0.91&15\\
              &0.57&0.04&-3.48&0.66&0.30&15\\
    \hline
    \pfh      &    &    &3.79 &0.15&1.14&13\\
              &0.61&0.05&-5.15&0.72&0.60&13\\
    \hline	   			   
    All       &    &    &41.53&1.09&0.73&12\\
              &2.84&0.34&-2.00&5.34&0.23&12\\
    \hline
    Pfund     &    &    &25.51&0.77&0.82&12\\
              &2.28&0.24&-9.52&3.79&0.23&12\\
    \hline 
    L'-band    &    &    &15.34&0.38&0.79&14\\
              &1.05&0.11&-0.40&1.71&0.43&14\\
    \hline	   			   
    \end{tabular}			   
    \label{dwarfs}			   
    \end{table}			   
    				       
    In the B-type dwarfs and giants, all hydrogen lines are seen in absorption,
    their strengths increasing with later spectral type.    
    This is most pronounced for the lowest Pfund series line observed (\pfg\ ), and 
    is less so for higher Pfund series lines and Brackett series lines (Table~\ref{dwarfs}).   
    This behaviour suggests that these lines might provide a spectral-type, i.e. temperature diagnostic. 
    All hydrogen lines show a similar first-order dependence, however, the
    slope for the Brackett lines is smaller than for the Pfund lines.
 
    The most accurate diagnostic for determining the spectral type from the equivalent widths
    of these lines is to add a number of equivalent widths.
    Adding all the lines we measured gives the stronger slope, but not the best relation
    to recover spectral types. Indeed adding the EW of the Pfund lines only, gives the same 
    measure of goodness of fit with smaller errors on the measurements. It is therefore a preferred 
    diagnostic. 

    We add the EW of the four Pfund lines we measured. The best linear fit 
    relation between spectral type and the summed EW is given in Table~\ref{dwarfs}.
    Using this relation, we are able to recover the spectral types of all twelve B dwarfs to giants
    used to define the fit, within two spectral sub-types. 
    Among those, for 8 of the 12 stars we find the spectral type to within 
    one sub-type, and for 6 of the 12 we recover the exact spectral type. 
    This result is quite satisfactory, considering that
    we adopted a simple linear fit to describe the EW versus spectral-type relation.
    				       
    The presence of \hei\ lines allows some refinement of our spectral-type
    estimates, as these lines appear only between spectral type O9 and B2 
    in dwarfs to giants. This allows us to assign $\alpha$~Pav, which was assigned type B4
    considering only the hydrogen lines, its correct spectral type: B2. 
    				       
    Using the summed EW of \bra\ and
    \pfg\ allows for a linear relation to determine the spectral type,
    identical to the method described previously. The parameters of
    this relation are also given in Table~\ref{dwarfs}. The linear fit recovers the spectral
    type of the 14 B-dwarfs and giants to within five spectral sub-types. 
    Of the 14, for 11 the classification is accurate to within four sub-types;
    for 10 it is within two sub-types; for 6 it is within one subtype, and for three
    it is exact. 
    At the extrema of the B classification, B0 and B9, the classification
    fails by five spectral sub-types, indicating earlier and later spectral types
    respectively. This suggests one must use a higher order fit
    and/or one has to separate the spectral-type dependence of dwarfs, sub-giants and
    giants. Unfortunately, the data quality and sample size does not allow us to
    investigate this possibility. We note that the L'-band spectral range 
    between 3.5 and 4.1~$\mu$m also contains
    some Humphreys series lines. However, these could not be used as their strength
    can only be accurately measured in late B-type stars.	   
    				       
    Given the data quality and spectral coverage of our sample, it is not possible
    to distinguish between giants and dwarfs using the equivalent widths only.
    However, the full width at half maximum (FWHM) of the \bra\ line does
    allow giants and dwarfs to be separated. 
    B-type dwarfs have a FWHM of more than 430 \kmsec\ (up to 665 \kmsec\ ), 
    after correction for the instrumental profile, and giants have a FWHM 
    of between 330 and 430 \kmsec. Supergiants
    that show a photospheric profile have even narrower \bra\ lines.
    The reason why a simple equivalent 
    width measurement fails to achieve this distinction can be explained by the quality of our
    data. Indeed, the main source of error in measuring the EW is in the position of 
    the continuum. Assuming Gaussian line shapes, and given our spectral resolution,
    the relative error in the EW is up to 2.5 times the relative error in the FWHM.
    We also tried to separate giants and dwarfs using the FWHM of \brb\ and \pfg,
    however, unfortunately without success. 
\subsection{O stars}
\label{Ostars}

     Simple relations connecting line strength to spectral type, such as
     for B dwarfs and giants (see Sect.~\ref{Bdwarfs}), cannot be derived for O-type 
     stars. The reason is a too limited sample of stars that is
     only observed in the L'-band. The difference in behaviour between
     Pf$\gamma$ and Br$\alpha$ also shows that mass loss plays an
     important role in the line formation process. Pf$\gamma$ shows a modest
     dependence of EW on spectral class -- dominated by temperature effects, 
     while Br$\alpha$ shows a steep dependence -- dominated by wind
     density effects. In the remainder of this section, we will 
     concentrate on the latter line as a diagnostic for stellar 
     mass loss $\dot{M}$. 
  
\begin{figure}[!ht]
     \resizebox{8.5cm}{!}{\includegraphics{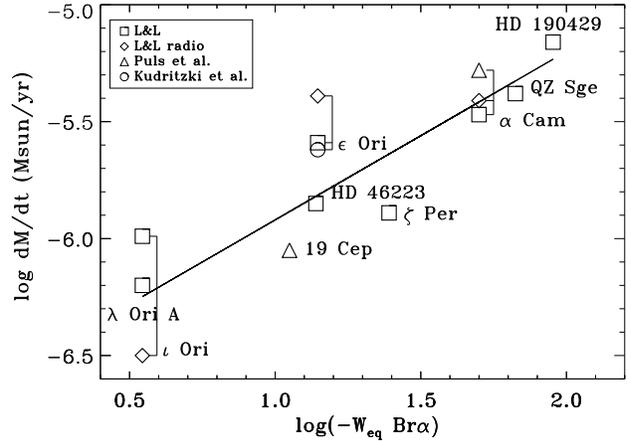}}
     \caption{The measured equivalent width of \bra\ vs. mass-loss 
              rate as determined from H$\alpha$ profile fitting and 
              radio measurements. For three stars multiple $\dot{M}$ 
              values are plotted, illustrating the intrinsic 
              uncertainty of mass-loss determination to be $\sim$ 
              0.2--0.3 dex. The adopted values for $\dot{M}$ are
              from Lamers \& Leitherer (1993); Puls et al. (1996), and
              Kudritzki et al. (1999). All rates are based on H$\alpha$
              fitting, except for three radio determinations (indicated
              by diamond symbols). See text for a discussion.
             }
     \label{fig:bra-mdot}
  \end{figure}

  All O stars in the sample show emission in \bra, except for
  two late-type main-sequence stars, i.e. HD\,47839 (O7\,V) 
  and HD\,37468 (O9.5\,V). The emission results from the 
  presence of strong stellar winds in these stars (see e.g.
  Kudritzki \& Puls 2000 for a review). This is illustrated
  in Fig.~\ref{fig:bra-mdot}, where the measured \bra\ equivalent
  width is plotted versus mass-loss rate. 
  For late O-type stars the \bra\ equivalent width 
  includes a non-negligible contribution of 
  He~{\sc i}\,$\lambda 4.4049$. The $\dot{M}$ values
  have been determined using either the strength of the H$\alpha$ 
  profile as a diagnostic or using radio fluxes. Most values are 
  from a compilation by Lamers \& Leitherer (1993). 
  Their H$\alpha$ rates are indicated by square symbols, while diamonds denote
  radio rates. Three additional measurements (from Puls et al.
  1996 and Kudritzki et al. 1999) are 
  based on fitting of the H$\alpha$ line. For three stars 
  ($\iota$\,Ori, $\epsilon$\,Ori, and $\alpha$\,Cam) multiple 
  mass-loss rate determinations are available. Intrinsic uncertainties
  in these determinations are typically 0.2 -- 0.3 dex, which
  is also illustrated by the range in values found for the
  three stars. The rather large difference in derived mass-loss rate
  for $\iota$\,Ori ($\dot{M}({\rm H}\alpha)$ = 10.2 \,10$^{-7}$
  vs. $\dot{M}({\rm radio}) = 3.2\,10^{-7}$ \msunyr) is likely
  related to the greater uncertainty in the treatment of the
  H$\alpha$ photospheric absorption as well as to the low flux 
  densities at cm wavelengths, for low values of $\dot{M}$.
  A clear relation between mass-loss rate and \bra\ equivalent width
  is present. Adopting an error of 0.2 (0.3) dex in the radio (H$\alpha$)
  rates and applying a weight average for the three stars for which multiple $\dot{M}$
  determinations are available, one finds a best fit linear relation:
  \begin{equation}
     \log \dot{M} =  (0.72 \pm 0.21) \log (-W_{\rm eq} Br\alpha) - (6.64 \pm 0.28)
     \label{eq:mdot-bra}
  \end{equation}
  
  The most accurate prediction for the mass-loss rate from an equivalent-width measurement
  is expected if the observed $W_{\rm eq}$ is corrected for photospheric
  absorption and is plotted versus the equivalent width invariant 
  $Q \equiv \dot{M}^{2}/(R^{3/2} \teff^{2}$ \vinf), which is essentially
  related to the wind density (de Koter et al. 1998, Puls et al. 
  1996). 
  This method requires accurate basic stellar parameters,
  which, as pointed out in this paper, are non-trivial to obtain if only 
  infrared data is available.
  Also, the terminal velocity \vinf\ of the stellar wind needs to be known.
  In O-type stars, this latter quantity is accurately 
  determined from the blue-edge in P~Cygni profiles of UV resonance 
  lines. If only the near-IR spectrum is observed, such a simple
  and accurate diagnostic to measure \vinf\ is not available. 
  For these reasons, we have
  opted to provide Eq.~\ref{eq:mdot-bra} as a simple means to obtain an
  estimate of the stellar mass-loss rate.
   We note that the relation can equally well be applied to B supergiants, 
   as both for O- and B-type stars the formation of H$\alpha$ is dominated by the recombination mechanism.

\section{B stars with emission lines}
\label{emmission}

In this section, we discuss the B stars with emission lines of our sample, this 
includes ``classical'' Be stars as well as B[e] stars and Luminous Blue 
Variable stars. Some of the spectra presented here have already been studied in 
great detail, e.g. $\gamma$~Cas in Hony et al. 2000.

In B stars with emission lines, most hydrogen lines are in emission 
in the 2.4-4.1 \mum\ range. Those emission lines mainly originate from 
circumstellar material that are filling in (partially or completely) 
the atmospheric absorption lines. 
The nature of the circumstellar material surrounding the objects of this sample
is very diverse. In Luminous Blue Variable stars, the emission lines originate
from a dense wind. B[e] stars (see Lamers et al. 1999 for a review) have 
(sometimes strong) forbidden lines implying that 
there is a large volume of low-density gas near the star in which conditions 
are favourable for the excitation of these transitions.
 
It is now well established that "classical" Be stars are surrounded by
dense, roughly keplerian circumstellar disks.  The most convincing
evidence for the presence of disks is derived from direct imaging at
optical wavelengths (e.g.  Quirrenbach et al. 1997) and at radio wavelengths
(Dougherty et al. 1992).  Besides imaging, other observed properties of Be stars
are also naturally explained by the presence of a circumstellar disk. 
One of the defining characteristics of Be stars is the presence of
(often double-peaked) H$\alpha$ emission.  The width of the H$\alpha$
line scales with the projected rotational velocity of the photosphere
($v \sin i$) (e.g. Dachs et al. 1986).  Both the double-peaked nature and the
relation between width and $v \sin i$ are consistent with the line emission
being formed in a flattened, rotating disk surrounding the star
(Poeckert et al. 1978).  In addition, the variations in the violet and red
peaks of the H$\alpha$ and other \hi\ lines in the spectra of Be stars are
explained due to spiral density waves in a non-self-gravitating
keplerian disk (Telting et al. 1994).  Such a keplerian disk geometry also
explains the continuum linear polarisation caused by Thomson scattering
of free electrons in the disk (e.g.  Cote et al. 1987).  The position angle
of the polarisation is consistent with the orientation of the disk
observed by imaging.  Be star disks tend to have large densities, as derived 
from e.g.  infrared excess (Waters et al. 1987). The disk radii probably vary 
from a fraction of a stellar radius (Cot\'e et al. 1996) to many tens 
of R$_{*}$ (Waters et al. 1991). 

We find no obvious correlation between spectral type and strength of the emission 
lines in the Be stars with luminosity class III to V (sometimes referred to as 
classical Be stars) in our sample (Fig.~\ref{Be1} and ~\ref{Be2}).
Other studies report a similar lack of correlation between spectral type and 
amount of circumstellar gas, except perhaps when it comes to the \emph{maximum}
amount of emission at a given spectral type (see e.g. Dougherty et al. 1992 or
Waters et al. 1986).
We do not see double peaked lines at our resolution ($\delta \lambda / \lambda 
\approx 1200$). 
We also do not find evidence for forbidden line emission in the
"classical" Be stars, in agreement with such a lack in optical
spectra. Further investigation is needed to conclude about the
presence of such lines in the spectra of the few B[e] of
our sample.  

\hei\ emission lines are present in most stars with spectral type earlier than B3. 
We find a few \oi\ emission lines, the strongest ones being at \lam2.8935 
and \lam3.6617 in several classical Be stars of spectral type earlier than B3 as
well as in Luminous Blue Variable stars and in the B[e] star HD~200775.
We also find \feii\ and \mgii\ emission lines in all three LBVs.
The \feii\ lines are also present in the spectra of HD~105435 and HD~45677.

The sample of B stars with emission lines will be investigated in more detail in
a forthcoming publication (Lenorzer et al. in prep.).


\section{Summary}
\label{summary}
  In this paper, we have presented an atlas of 2.4 to 4.1 \mum\ ISO SWS spectra of 
  early-type stars, mainly obtained during its Post-Helium mission, and several 3.5-4.1 \mum\ 
  spectra of O stars obtained at UKIRT.
  The observations include normal OB, Be and Luminous Blue Variable stars.
  Later spectral types will be presented in a separate publication (Vandenbussche et al. 
  in prep.). We have explored a number of simple empirical methods aimed at using the 
  infrared spectrum to {\em i)} determine the spectral type and/or luminosity class, 
  and {\em ii)} determine the mass-loss rate. The main results are:
  		 
  \begin{enumerate}
  		 
  \item In normal B-type giant to dwarf stars the Pfund lines, and to a lesser extent 
  	the Brackett lines, may be used to estimate the spectral type. 
        We provide a simple formula to do this.  
        The leading line of
  	each series shows the most pronounced dependence. Helium lines help to
  	improve this spectral classification, \hei\ being present in late O-type and 
        early B-type stars. 
        All B-type giants and dwarfs have \bra\ in absorption.
        The full width at half maximum of this line may be used to
        discriminate between luminosity classes III and V, the line
        being broader for dwarfs. \\
  		 
  \item  In B-type supergiants the equivalent width of all measured hydrogen
         lines remain constant with spectral sub-type, although with a
         significant scatter. \bra\ is seen mostly in emission, while
         all other lines are in absorption. \hei~$\lambda$3.0736
         is systematically stronger in absorption compared to B-type
         dwarfs and giants.\\

  \item In normal O-type stars and in B-type supergiants, the \bra\ line is mostly 
  	in emission and provides a sensitive indicator of the mass-loss rate. 
        We give a relation that uses the equivalent width of this line to estimate
        $\dot{M}$.\\

  \item Concerning hydrogen lines, the ones positioned in the L'-band
         seem best suited to derive physical properties of OB stars when
         compared to the diagnostics available in other atmospheric
         bands such as K-, H-, and J-band. The main reason is that the
         L'-band contains three different hydrogen series lines and
         includes the leading Brackett-series line. Concerning other
         species, the K-band seems to contain the most useful lines. 
         This last remark, however, only applies to O-type stars (where 
         e.g. C~{\sc iv}, N~{\sc iii} and an unblended He~{\sc ii} line 
         are seen) and not to B-type stars which do not show lines of 
         metal species in that wavelength range.\\

 \item  In our sample of Be, B[e], and Luminous Blue Variable stars
         we find no obvious correlation between spectral type and strength
         of the emission lines. Stars with spectral type earlier
         than B3 show \hei\ lines, similar to normal B-type stars.
         Several emission line stars show \oi, however not at 
         spectral types later than B2.\\

\end{enumerate} 
  	
\begin{acknowledgements}
We thank Jan Cami for stimulating discussions and help in data
processing. This work was supported by NWO Pionier grant 600-78-333. AdK kindly
acknowledges support from NWO Spinoza grant 08-0 to E.P.J. van den Heuvel.
LK is supported by a fellowship of the Royal Academy of Sciences in the 
Netherlands. TRG is supported by the Gemini Observatory, which is operated 
by the Association of Universities for Research in Astronomy, Inc., on 
behalf of the international Gemini parthership of Argentina, Australia, 
Brazil, Canada, Chile, the United Kingdom and the United States of America. 
We acknowledge the use of the Atomic Line List compiled by Peter van Hoof, 
which can be accessed through the web at http://www.pa.uky.edu/~peter/atom
ic/index.html.
\end{acknowledgements}

\appendix
\bf APPENDIX : Equivalent width measurements
\end{document}